\documentstyle[aps,psfig]{revtex}

\begin{document}

\title{Statistical Properties of Interacting Bose Gases in Quasi-2D Harmonic 
       Traps}

\author{Hongwei Xiong$^{1,2}$, Shujuan Liu$^{1}$, Guoxiang Huang$^{3}$}
\address{$^{1}$Department of Applied Physics, Zhejiang
University of Technology, Hangzhou, 310032, China}
\address{$^{2}$Zhijiang College,
Zhejiang University of Technology, Hangzhou, 310012, China }
\address{$^{3}$Department of Physics, East China Normal University, Shanghai, 200062, 
China}
\date{\today}
\maketitle

\begin{abstract}
{\it The analytical probability distribution of the quasi-2D (and purely 2D) 
ideal and interacting Bose gas are investigated by using a canonical 
ensemble approach. Using the analytical probability 
distribution of the condensate, the statistical properties such as 
the mean occupation number and particle number fluctuations of the 
condensate are calculated. Researches show that there 
is a continuous crossover of the statistical properties from a 
quasi-2D to a purely 2D ideal or interacting gases. Different from the 
case of a 3D Bose gas, the interaction between atoms
changes in a deep way the nature of the particle number fluctuations.\\}
hongweixiong@hotmail.com\\
pacs: 03.75.Fi, 05.30.Jp
\end{abstract}


\section{Introduction}


The remarkable observations of Bose-Einstein condensation (BEC) in atomic
vapors \cite{ALK,MIT,HEL} have opened up new avenues of research into the
physical properties and nature of Bose-condensed systems \cite
{RMP,PARKIN,LEG}. Recently BECs have been realized in quasi-one and
quasi-two dimensions \cite{EXPLOW,LOW2}, where new phenomena such as
quasicondensates with a fluctuating phase \cite{PETROV1,PETROV2,Kagan} and a
Tonks gas of impenetrable bosons \cite{PETROV2,Tonks,Olshanii} may be
observed. After the realization of BEC, the statistical properties of the
system (especially the particle number fluctuations of the condensate) have
been studied in deep for three-dimensional (3D) ideal Bose gases confined in
a box \cite{BOR,WIL}, and in the presence of a harmonic trap \cite
{WIL,POL,GAJ,GRO,NAVEZ,BAL,HOL1,HOL2}. Recently the question of how
interatomic interactions affect the particle number fluctuations has also
been the object of several theoretical investigations \cite
{GIO,IDZ,ILLU,MEI,KOC,JAK,XIONG1,XIONG2}. Recall that BECs can not occur in
one-dimensional (1D) and two-dimensional (2D) uniform Bose gases at finite
temperature, because in that case thermal fluctuations would destabilize the
condensate. The realization of the trapped BECs in various dimensions makes
the behavior of particle number fluctuations a very interesting problem,
especially for 1D and 2D Bose gases in a harmonic trap.

It is well known that the theory based on a grand-canonical ensemble gives
unphysically large particle number fluctuations of condensate, in
particular when the temperature approaches zero. It is therefore
necessary to use a canonical (or microcanonical) ensemble to investigate the statistical
properties of the trapped Bose gas.
Recently a simple method has been developed \cite{XIONG1,XIONG2,XIONG3} to
calculate the probability distribution of the condensate within 
canonical ensembles. As soon as the probability distribution of the
condensate is known, the statistical properties of the system can be obtained
directly. With this method the mean ground state
occupation number and particle number fluctuations for the 3D 
have been considered for interacting
gas trapped in a box \cite{XIONG1} and a harmonic potential \cite{XIONG2}.
The role of dimensionality \cite{XIONG3} on the statistical properties of
the ideal gas was also discussed recently through the calculations of
the probability distribution for different spatial dimensions
and various confining potentials.
The purpose of the present work is an attempt to investigate the probability
distribution of quasi-2D (and purely 2D) ideal and interacting gases 
confined in a
harmonic trap. Using the probability distribution function obtained, 
we can calculate the statistical properties of the system such as the mean
occupation number and particle number fluctuations of condensate in a simple
way.

The paper is organized as follows. In Sec. II we give a simple introduction
of our approach to calculate the statistical properties of the Bose gas
within the canonical ensemble. In Secs. III and IV we investigate the
statistical properties of quasi-2D (and purely 2D) ideal and interacting
Bose gases, respectively. The last section (Sec. V) contains a discussion and
summary of our results.


\section{Probability Distribution within Canonical Ensemble}


The statistical method used in this work follows from that used for the
3D dilute interacting gases developed in Ref. \cite{XIONG1,XIONG2},
where a canonical ensemble is used to discuss the statistical
properties of the system. We start from the canonical partition
function of $N$ trapped interacting bosons, given by

\begin{equation}
{Z}\left( N\right) {=\sum_{\Sigma _{{\bf {n}}}N_{{\bf {n}}}=N}\exp \left[
-\beta \left( E_{\bf{0}}+ \Sigma _{{\bf {n} \neq {\bf{0}} }}N_{{\bf {n}}}\varepsilon _{{\bf {n}}
}+E_{int}\right) \right] },  \label{par1}
\end{equation}

\noindent where $\beta =1/k_{B}T$, $N_{{\bf n}}$ and $\varepsilon _{{\bf n}}$
are occupation number and energy level of the state ${\bf {n}}$,
respectively. $E_{\bf 0}$ is the energy of the condensate of the system. In the case 
of non-interacting Bose gas, $E_{\bf 0}=N_{\bf 0}\varepsilon _{\bf 0}$; for the case of
interacting Bose gas, however, $E_{\bf 0}$ should be regarded as the interaction energy
of the condensate. In the above equation, 
$E_{int}$ is the interaction energy between the condensate and the normal gas, which
takes the form

\begin{equation}
E_{int}=2g\int n_{0}\left(\bf r\right) n_{T}\left(\bf r\right) d^{3}{\bf r}+g\int n_{T}^{2}\left(\bf r\right) d^{3}{\bf r},
\label{interinterction}
\end{equation}
where $n_{0}\left(\bf r\right)$ and $n_{T}\left(\bf r\right)$ are the density distribution of the condensate
and normal gas, respectively. In the above equation, $g$ represents the coupling constant.

The probability distribution of a condensate is defined as the probability
to find $N_{{\bf {0}}}$ atoms in the condensate. Through a developed
saddle-point method \cite{XIONG1,XIONG2} one can obtain the probability
distribution of the condensate, which is found to be

\begin{equation}
{G}_{n}{\left( N,N_{{\bf {0}}}\right) =A}_{n}{\exp \left[ \int_{N_{{\bf {0}}
}^{p}}^{N_{{\bf {0}}}}\alpha \left( N,N_{{\bf {0}}}\right) dN_{{\bf {0}}
}\right] ,}  \label{disfunction}
\end{equation}
where ${A}_{n}$ and $N_{{\bf {0}}}^{p}$ are normalization constant and the
most probability value of the condensate, respectively. The quantity
$\alpha \left( N,N_{
{\bf {0}}}\right) $ can be obtained based on the following equation:

\begin{equation}
{N_{{\bf 0}}=N-\sum_{{\bf n\neq 0}}\frac{1}{\exp \left[ \beta \varepsilon _{
{\bf n}}\right] \exp \lbrack -\beta {\frac{\partial }{\partial N_{{\bf {0}}}}
\left( E_{{\bf {0}}}+E_{int}\right) -\alpha \left( N,N_{{\bf {0}}}\right) }
\rbrack -1}.}  \label{alphafunction}
\end{equation}
By setting ${\alpha \left( N,N_{{\bf {0}}}\right) =0}$ in Eq. (\ref{alphafunction})
one gets the most probability value $N_{{\bf {0}}}^{p}$,

\begin{equation}
{N_{{\bf {0}}}^{p}=N-\sum_{{\bf n}\neq {\bf 0}}\frac{1}{\exp \left[ \beta
\varepsilon _{{\bf n}}\right] \exp \lbrack -\beta {\frac{\partial }{\partial
N_{{\bf {0}}}^{p}}\left( E_{{\bf {0}}}+E_{int}\right) }\rbrack -1}.}
\label{mostprobability}
\end{equation}
Note that ${N_{{\bf {0}}}^{p}}$ is exactly the mean ground state occupation
number within a grand canonical ensemble, when the lowest order
perturbation theory is used to discuss the role of atom-atom interactions
(see Ref. \cite{XIONG1,XIONG2,XIONG4}). Although ${\alpha \left( N,N_{{\bf {0
}}}\right) }$ can be obtained directly from Eq. (\ref{alphafunction}), it is
more convenient to calculate it through the difference between ${N_{{\bf 0}}}
$ and ${N_{{\bf {0}}}^{p}},$ {\it i.e.},

\[
{N_{{\bf {0}}}\vspace{1pt}-N_{{\bf {0}}}^{p}=\sum_{{\bf n}\neq {\bf 0}}\frac{
1}{\exp \left[ \beta \varepsilon _{{\bf n}}\right] \exp \lbrack -\beta {
\frac{\partial }{\partial N_{{\bf {0}}}^{p}}\left( E_{{\bf {0}}
}+E_{int}\right) }\rbrack -1}} 
\]

\begin{equation}
{-\sum_{{\bf n}\neq {\bf 0}}\frac{1}{\exp \left[ \beta \varepsilon _{{\bf n}
}\right] \exp \lbrack -\beta {\frac{\partial }{\partial N_{{\bf {0}}}}\left(
E_{{\bf {0}}}+E_{int}\right) -\alpha \left( N,N_{{\bf {0}}}\right) }\rbrack
-1}.}  \label{alphao}
\end{equation}
Once we know $\varepsilon _{{\bf n}}$, $E_{{\bf {0}}}$ and $E_{int}$ of the
system, it is straightforward to obtain $\alpha \left( N,N_{{\bf {0}}
}\right) $ from Eq. (\ref{alphao}). Then by Eq. (\ref{disfunction})
one can obtain the probability distribution of the condensate directly.

As soon as we know $G_{n}\left( N,N_{{\bf {0}}}\right) $, the statistical
properties of the system can be clearly described. From the probability
distribution function given by Eq. (\ref{disfunction}) one obtains $
\left\langle N_{{\bf {0}}}\right\rangle $ and $\left\langle \delta ^{2}N_{
{\bf {0}}}\right\rangle $ within the canonical ensemble:

\begin{equation}
{\left\langle N_{{\bf {0}}}\right\rangle =\frac{\sum_{N_{{\bf {0}}}=0}^{N}N_{
{\bf {0}}}G_{n}\left( N,N_{{\bf {0}}}\right) }{\sum_{N_{{\bf {0}}%
}=0}^{N}G_{n}\left( N,N_{{\bf {0}}}\right) },}  \label{mean}
\end{equation}

\begin{equation}
{\left\langle \delta ^{2}N_{{\bf {0}}}\right\rangle =\frac{\sum_{N_{{\bf {0}}
}=0}^{N}N_{{\bf {0}}}^{2}G_{n}\left( N,N_{{\bf {0}}}\right) }{\sum_{N_{{\bf {
0}}}=0}^{N}G_{n}\left( N,N_{{\bf {0}}}\right) }-\left[ \frac{\sum_{N_{{\bf {0
}}}=0}^{N}N_{{\bf {0}}}G_{n}\left( N,N_{{\bf {0}}}\right) }{\sum_{N_{{\bf {0}
}}=0}^{N}G_{n}\left( N,N_{{\bf {0}}}\right) }\right] ^{2}.}  \label{fluc}
\end{equation}

\noindent We see that in this approach, the calculation of $\alpha \left(
N,N_{{\bf {0}}}\right) $ by using Eq. (\ref{alphao}) plays a crucial role to
discuss the statistical properties of the condensate.


\section{Ideal Bose Gases in a Quasi-2D Harmonic Trap}


Now we apply the formulation introduced in the last section to calculate the
thermodynamical quantities such as the particle number fluctuations of the
condensate for an ideal 2D Bose gas. We first consider the statistical
properties of a purely 2D Bose gas confined in axially symmetric traps. 
The purely 2D Bose gas can be regarded as the limit of a trapped 3D Bose gas
with the trapping frequency ${\omega _{z}}$
in the $z$ direction approaching infinity.

The single-particle energy level of the purely 2D Bose gas in an axially
symmetric harmonic trap takes the form

\begin{equation}
{\varepsilon _{{\bf {n}}}=\left( n_{x}+n_{y}+1\right) \hbar \omega _{\perp },
}  \label{2D-energy}
\end{equation}
where ${\omega _{\perp }=\omega _{z}=\omega _{y}}$ is the trapping frequency
in the radial direction. From Eqs. (\ref{alphao}) and (\ref{2D-energy}), one
gets

\begin{equation}
{N_{{\bf 0}}-N_{{\bf 0}}^{p}=\sum_{{\bf n}\neq {\bf 0}}}\left[ {\frac{1}{
\exp \left[ \beta \left( n_{x}+n_{y}\right) \hbar \omega _{\perp }\right] -1}
}-{\frac{1}{\exp \left[ \beta \left( n_{x}+n_{y}\right) \hbar \omega _{\perp
}\right] \exp \lbrack -{\alpha \left( N,N_{{\bf {0}}}\right) }\rbrack -1}}
\right] {.}  \label{2Dalpha}
\end{equation}

Assuming $\left| {\alpha \left( N,N_{{\bf {0}}}\right) }\right| <<1$, 
we obtain 
the result of ${\alpha \left( N,N_{{\bf {0}}}\right) }$:

\begin{equation}
{\alpha \left( N,N_{{\bf {0}}}\right) =-}\frac{{N_{{\bf 0}}-N_{{\bf 0}}^{p}}
}{\Xi _{2D}},  \label{2Dalpha2}
\end{equation}
where

\begin{equation}
\Xi _{2D}=\sum_{{\bf n}\neq {\bf 0}}\frac{\exp \left[ \beta \left(
n_{x}+n_{y}\right) \hbar \omega _{\perp }\right] }{\left\{ \exp \left[ \beta
\left( n_{x}+n_{y}\right) \hbar \omega _{\perp }\right] -1\right\} ^{2}}.
\label{2Dxi}
\end{equation}

\noindent It is easy to find that Eq. (\ref{2Dalpha2})
coincides with the assumption $\left| {\alpha \left( N,N_{{\bf {0}}}\right) }
\right| <<1$. Using Eq. (\ref{disfunction}), the probability distribution of
the purely 2D ideal gas is then given by

\begin{equation}
G_{2D}=A_{2D}\exp \left[ -\frac{\left( {N_{{\bf 0}}-N_{{\bf 0}}^{p}}\right)
^{2}}{2\Xi _{2D}}\right] ,  \label{fluc2d}
\end{equation}
where $A_{2D}$ is a normalization constant.

In addition, from Eqs. (\ref{mostprobability}) and (\ref{2D-energy}), we obtain the most probable
value ${N_{{\bf 0}}^{p}}$ as \cite{MULLIN}:

\begin{equation}
{N_{{\bf 0}}^{p}=N-}Nt^{2}-{tN}^{1/2}\ln N,  \label{2D-Np}
\end{equation}
where we have introduced a reduced temperature $t=T/T_{c}^{0}$, with 
$T_{c}^{0}=\left( N/\zeta \left( 2\right)\right) ^{1/2}\hbar
\omega _{\perp }/k_{B}$ being the critical temperature of the system in the
thermodynamic limit.

Using the formulas (\ref{mean}), (\ref{fluc}), (\ref{fluc2d}) and
(\ref{2D-Np}), we can obtain the mean ground state occupation number and
particle number fluctuations of the condensate. In particular, below the
critical temperature ({\it i.e.}, in the case of ${N_{{\bf 0}}^{p}>>1}$), the
analytical result of the particle number fluctuations is given by

\begin{equation}
{\left\langle \delta ^{2}N_{{\bf 0}}\right\rangle }_{2D}{=}\Xi _{2D}\approx 
\frac{Nt^{2}}{\zeta \left( 2\right) }\left[ \frac{1}{2}\ln \left( \frac{
Nt^{2}}{\zeta \left( 2\right) }\right) +1\right].  \label{2danalytical}
\end{equation}
When obtaining Eq. (\ref{2danalytical}) we have used the formula 
${\left\langle \delta ^{2}N_{{\bf 0}}\right\rangle =}a/2$ for {the particle
number fluctuations} when the probability distribution is a Gaussian
function with the form 
$G_{n}=A_{n}\exp \left[ -\left( {N_{{\bf 0}}^{p}-N_{{\bf 0}}}
\right) ^{2}/a\right] $, and ${N_{{\bf 0}}^{p}>>1}$. A simple
derivation for this result has been given in the Appendix of 
Ref. \cite{XIONG3}.

We see that the particle number fluctuations of the harmonically trapped 2D
gas exhibit a very weakly anomalous behavior, as it is controlled by the
factor $\ln \left( Nt^{2}/\zeta \left( 2\right) \right) $. Shown in Fig. 1
is the numerical result of $\delta N_{{\bf 0}}=\sqrt{{\left\langle \delta
^{2}N_{{\bf 0}}\right\rangle }_{2D}}$ (solid line) for $N=10^{3}$ purely 2D
ideal bosons confined in an axially symmetric trap. The numerical result of $
{\left\langle N_{{\bf 0}}\right\rangle /N}$ (solid line) for $N=10^{3}$ purely 2D
ideal bosons is demonstrated in Fig. 2. As a comparison, 
${\left\langle N_{{\bf 0}}\right\rangle /N}$ within a grand-canonical ensemble
(or ${N_{{\bf 0}}^{p}}$ within the canonical ensemble) is also shown in the figure. It is
easy to show that ${N_{{\bf 0}}^{p}}$ coincides with ${\left\langle
N_{{\bf 0}}\right\rangle }$ in the case of ${N_{{\bf 0}}^{p}>>1}$. However,
there is an obvious difference between ${N_{{\bf 0}}^{p}}$ and ${
\left\langle N_{{\bf 0}}\right\rangle }$ near the critical temperature when
the total number of particles $N$ is smaller than $5\times 10^{3}$.

\begin{figure}[tb]
\psfig{figure=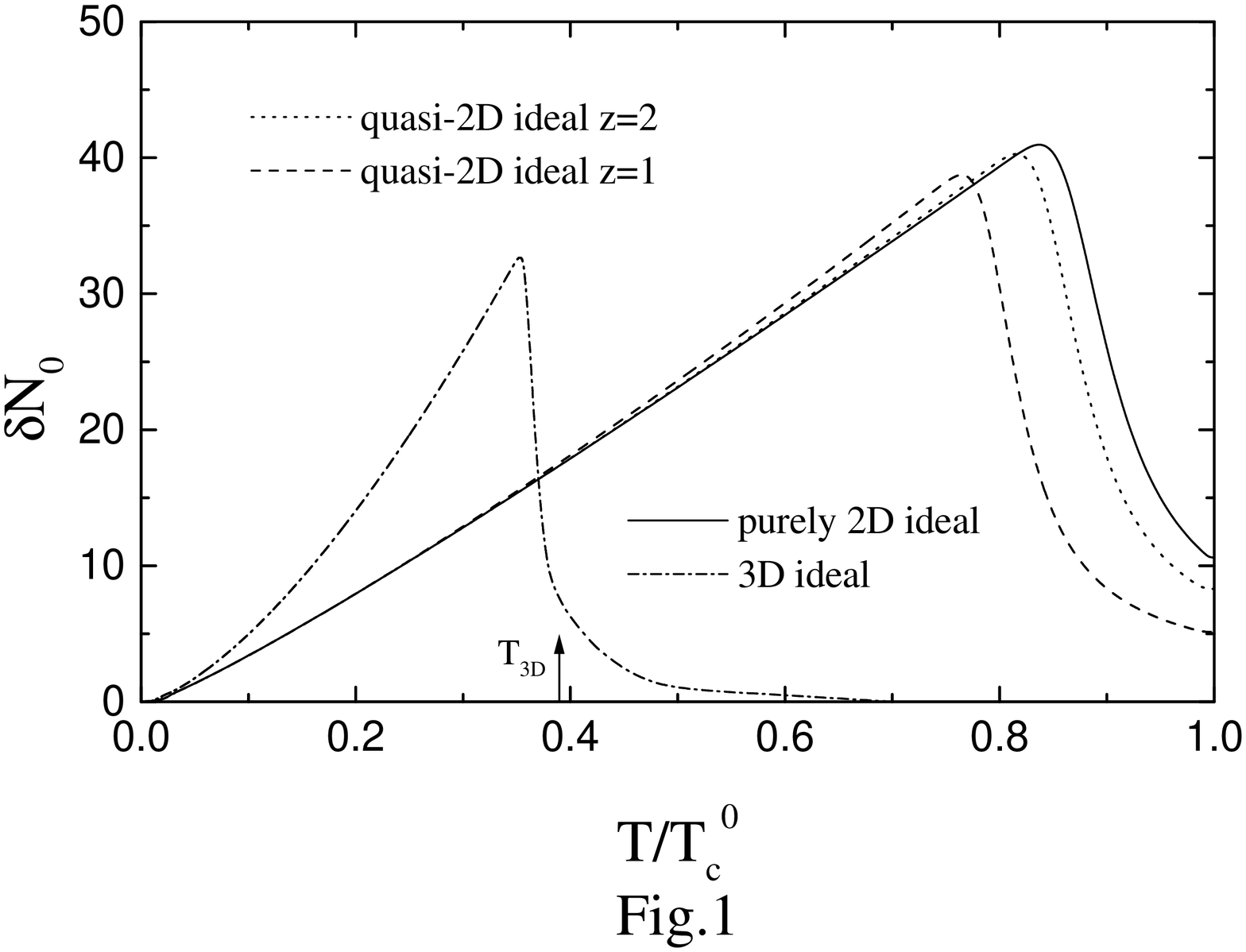,width=0.5\columnwidth,angle=270}
\caption{Root-mean-square fluctuations $\delta N_{{\bf {0}}}$ versus $
T/T_{c}^{0}$ for $N=10^{3}$ non-interacting bosons confined in purely 2D and
quasi-2D axially symmetric traps. The solid line displays the numerical
result of $\delta N_{{\bf {0}}}$ in a purely 2D harmonic trap, while the
dashed and dotted lines show the result of a quasi-2D trap with $z=1$ and $z=2$,
respectively (Here, $z$ is defined to be $z=\hbar \omega _{z}/k_{B}T_{c}^{0}$). 
The crossover of $\delta N_{{\bf 0}}$ from quasi-2D
to purely 2D gases is clearly illustrated in the figure. As illustrated in
the figure, below the critical temperature, $\delta N_{{\bf 0}}$ of the quasi-2D gas is slightly larger than
the result of the purely 2D gas because confinement has the effect of
reducing the particle number fluctuations.
}
\end{figure}

\begin{figure}[tb]
\psfig{figure=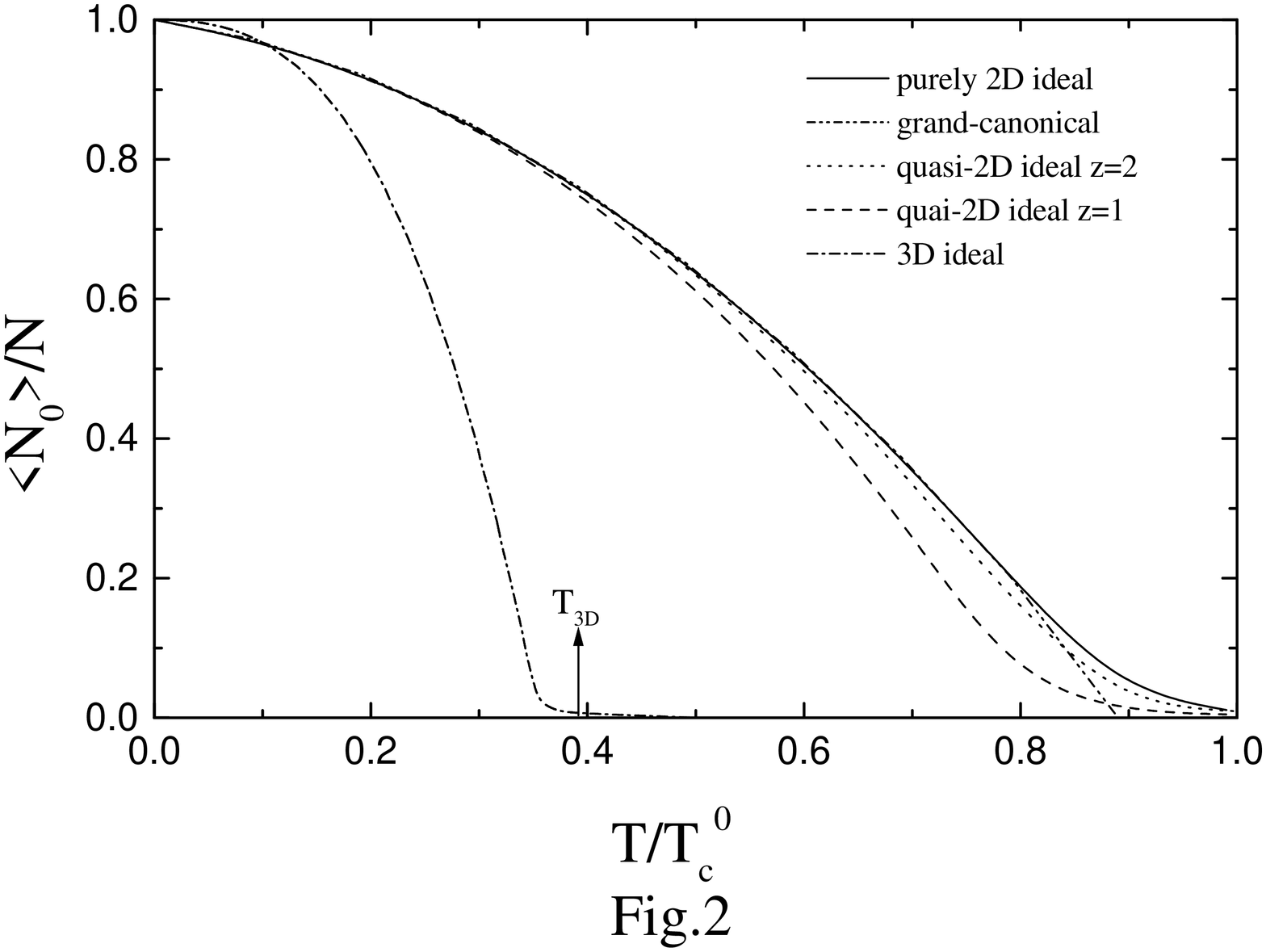,width=0.5\columnwidth,angle=270}
\caption{Displayed is the numerical result of the relative mean ground state
occupation number ${\left\langle N_{{\bf 0}}\right\rangle /N}$ for $N=10^{3}$
purely 2D (solid line) and quasi-2D ideal bosons. 
${\left\langle N_{{\bf 0}}\right\rangle /N}$ within the grand-canonical ensemble 
(or ${N_{{\bf 0}}^{p}/N}$ within the canonical ensemble) is also shown for $N=10^{3}$
purely 2D ideal bosons. We see in the figure that ${\left\langle N_{{\bf 0}
}\right\rangle /N}$ in the quasi-2D gas is slightly smaller than the result
of the purely 2D gas.
}
\end{figure}

In the recent experiment \cite{EXPLOW}, a quasi-2D BEC has been realized by
choosing the frequency $\omega _{z}$ large enough in a 3D trap. For a
real 3D axially symmetric trap, the single-particle energy level is ${
\varepsilon _{{\bf {n}}}=\left( n_{x}+n_{y}+1\right) \hbar \omega _{\perp
}+(n}_{z}+1/2{)\hbar \omega _{z}}$. The occupation number $N_{{\bf n}}$
of the exited ${\bf n\neq 0}$ is then given by

\begin{equation}
N_{{\bf n}}={\frac{1}{\exp \left[ \beta {\left( n_{x}+n_{y}\right) \hbar
\omega _{\perp }}\right] \exp \lbrack \beta {n}_{z}{\hbar \omega _{z}}
\rbrack -1}.}
\end{equation}
We see that in the case of ${\hbar \omega _{z}>}k_{B}T$ the occupation
number of the excited state $n_{z}\neq 0$ is very small, {\it i.e.}, the
oscillation motion is frozen in the $z$ direction. Thus, for an ideal Bose
gas, in order to achieve 2D Bose-Einstein condensation in a real 3D trap, we
should choose the frequency $\omega _{z}$ large enough to satisfy the
condition $\hbar \omega _{\perp }<<k_{B}T<\hbar \omega _{z}$. The system is
said to be a quasi-2D one when this condition is satisfied. To display the property
of the quai-2D Bose gas, it is useful to define a dimensionless parameter
$z=\hbar \omega _{z}/k_{B}T_{c}^{0}$, which represents the ratio between 
$\hbar \omega _{z}$ and $k_{B}T_{c}^{0}$.

From Eq. (\ref{alphafunction}), for the quasi-2D ideal gas we get

\[
{{N_{{\bf 0}}=N-\sum_{n_{x},n_{y}\neq 0}^{\infty }}\left[ {\frac{1}{\exp
\left[ \beta \left( n_{x}+n_{y}\right) \hbar \omega _{\perp }\right] \exp
\lbrack -{\alpha \left( N,N_{{\bf {0}}}\right) }\rbrack -1}}\right] } 
\]

\begin{equation}
~~~~~~~~~~~~~-{\sum_{n_{z}=1}^{\infty }\sum_{n_{x},n_{y}=0}^{\infty
}}\left[ {\frac{1}{\exp \left[ \beta \left( n_{x}+n_{y}\right) \hbar \omega
_{\perp }\right] \exp \left[ \beta n_{z}\hbar \omega _{z}\right] \exp
\lbrack -{\alpha \left( N,N_{{\bf {0}}}\right) }\rbrack -1}}\right] .
\label{q2di}
\end{equation}
For convenience we separate off the state $n_{z}=0$ from the state
$n_z\neq 0$ in
the sum on the right hand side of Eq. (\ref{q2di}). By setting ${\alpha
\left( N,N_{{\bf {0}}}\right) =0}$ in Eq. (\ref{q2di}), we get 
${{N_{{\bf 0}}^{p}}}$:

\[
{{N_{{\bf 0}}^{p}=N-\sum_{n_{x},n_{y}\neq 0}^{\infty }}\left[ {\frac{1}{\exp
\left[ \beta \left( n_{x}+n_{y}\right) \hbar \omega _{\perp }\right] -1}}
\right] } 
\]

\begin{equation}
~~~~~~~~~~~~~~~~~~~~~~-{\sum_{n_{z}=1}^{\infty }\sum_{n_{x},n_{y}=0}^{\infty
}}\left[ {\frac{1}{\exp \left[ \beta \left( n_{x}+n_{y}\right) \hbar \omega
_{\perp }\right] \exp \left[ \beta n_{z}\hbar \omega _{z}\right] -1}}\right]
.  \label{q2dip}
\end{equation}

From Eqs. (\ref{q2di}) and (\ref{q2dip}), it is straightforward to get the
finial result of ${\alpha \left( N,N_{{\bf {0}}}\right) }$

\begin{equation}
{\alpha \left( N,N_{{\bf {0}}}\right) =-}\frac{{N_{{\bf 0}}-N_{{\bf 0}}^{p}}
}{\Xi _{2D}+\Theta _{2D}},  \label{qalpha}
\end{equation}
where $\Xi _{2D}$ is the constant determined by Eq. (\ref{2Dxi}), and

\begin{equation}
\Theta _{2D}={\sum_{n_{z}=1}^{\infty }\sum_{n_{x},n_{y}=0}^{\infty }}\frac{
\exp \left[ \beta \left( n_{x}+n_{y}\right) \hbar \omega _{\perp }\right]
\exp \left[ \beta n_{z}\hbar \omega _{z}\right] }{\left\{ \exp \left[ \beta
\left( n_{x}+n_{y}\right) \hbar \omega _{\perp }\right] \exp \left[ \beta
n_{z}\hbar \omega _{z}\right] -1\right\} ^{2}}.  \label{qtheta1}
\end{equation}
For the quasi-2D gas, we have $k_{B}T<\hbar \omega _{z}$. In this situation, $\Theta _{2D}$ can be
approximated as:

\begin{equation}
\Theta _{2D}=\frac{Nt^{2}}{\zeta \left( 2\right) }\left[ {
\sum_{n_{z}=1}^{\infty }\exp }\left( -\beta n_{z}\hbar \omega _{z}\right)
\right] .  \label{qtheta}
\end{equation}

The probability distribution of the quasi-2D gas is then

\begin{equation}
G_{q-2D}=A_{q-2D}\exp \left[ -\frac{\left( {N_{{\bf 0}}-N_{{\bf 0}}^{p}}
\right) ^{2}}{2\left( \Xi _{2D}+\Theta _{2D}\right) }\right] ,  \label{qdis}
\end{equation}
where $A_{q-2D}$ is a normalization constant. In addition, from Eq. (\ref
{q2dip}), the most probability value can be approximated as:

\begin{equation}
{N_{{\bf 0}}^{p}=N-}Nt^{2}-tN^{1/2}\ln N-\Theta _{2D}{.}  \label{q2dip1}
\end{equation}
The last term $\Theta _{2D}$ on the right hand side of \ Eq. (\ref{q2dip1})
represents a modification between a quasi-2D and a purely 2D Bose gas.

From the formulas (\ref{mean}), (\ref{fluc}), (\ref{qdis}) and (\ref{q2dip1}),
we can obtain the mean ground state occupation number and particle number
fluctuations of the condensate for the quasi-2D gas. Below the critical
temperature, the particle number fluctuations of the condensate for the
quasi-2D gas is given by

\begin{equation}
{\left\langle \delta ^{2}N_{{\bf 0}}\right\rangle }_{q-2D}=\Xi _{2D}+\Theta
_{2D}.  \label{q2dflu}
\end{equation}
Similar to the most probability value, the last term $\Theta _{2D}$ on
the right hand side of \ Eq. (\ref{q2dflu}) represents a modification
between the quasi-2D and the purely 2D Bose gas.

In Fig. 1, the numerical results of $\delta N_{{\bf 0}}$ for the quasi-2D ideal gas
are shown for $z=1$ (dashed line) and $z=2$ (dotted line), respectively.
The continuous crossover of $\delta N_{{\bf 0}}$ from the quasi-2D to the purely 2D
gases is clearly illustrated in the figure. Below the critical temperature, we see that
$\delta N_{{\bf 0}}$
of the quasi-2D gas is slightly larger than the result of the purely 2D gas
because confinement has the effect of reducing the particle number
fluctuations. As a comparison,
the dot-dashed line in Fig. 1 displays $\delta N_{{\bf 0}}$ of $N=10^{3}$ ideal 
bosons confined in a 3D istropic harmonic trap, where the oscillator frequency
is assumed to be equal to the transverse frequency $\omega _{\perp }$
of 2D Bose gas. The arrow marks the critical temperature of the 3D Bose gas
in the thermodynamic limit.
Shown in Fig. 2 are the numerical results of ${\left\langle N_{
{\bf 0}}\right\rangle /N}$ for the quasi-2D ideal gas with
$z=1$ (dashed line) and $z=2$ (dotted line). There is also a continuous crossover of 
${\left\langle N_{{\bf 0}}\right\rangle /N}$ from quasi-2D to purely 2D
gases. In addition, ${\left\langle N_{{\bf 0}}\right\rangle /N}$ for the
quasi-2D gas is slightly smaller than the result of the purely 2D gas.


\section{Interacting Bose Gases in a Quasi-2D Harmonic Trap}


In this section we turn to discuss the statistical properties of a quasi-2D
interacting Bose gas in an axially symmetric trap. The interaction between
atoms can be described by a single coupling constant $g\approx 2\sqrt{2\pi }
\hbar ^{2}a_s/\left( ml_{z}\right) $ \cite{PETROV1}\ which is fixed by a $s$
-wave scattering length $a_s$ and the oscillator length $l_{z}=(\hbar /m\omega
_{z})^{1/2}$ in the $z$ direction. Note that the coupling constant in the
quasi-2D interacting gas is related to the trapping frequency $\omega _{z}$
in the $z$ direction because it is the trapping frequency $\omega _{z}$ that
leads to the crossover from a 3D to a quasi-2D gas. This unique property
makes the statistical properties of the quasi-2D interacting gas a very
interesting problem. By the investigation of the statistical properties
of the quasi-2D interacting gas, one can get a clear description of the role
of the trapping frequency $\omega _{z}$. For a weakly repulsive interacting
Bose-condensed gas, the correlation length $\hbar /\sqrt{mn_{0}g}$ ($n_{0}$
is the density distribution of the condensate) should greatly exceed the
mean interparticle separation $1/\sqrt{2\pi n_{0}}$. In this situation, one
can introduce a small parameter $\theta =mg/2\pi \hbar ^{2}<<1$ for the weakly
interacting gas \cite{PETROV1}.

In Thomas-Fermi regime the chemical potential of the quasi-2D gas is
given by $\mu \left( N_{{\bf 0}},T\right) =(N_{{\bf 0}}mg/\pi )^{1/2}\omega
_{\perp }$.
In addition, there is an important relation between $E_{\bf 0}$ and
$\mu \left( N_{{\bf 0}},T\right)$: $\mu \left( N_{\bf 0},T\right) =\partial E_{\bf 0}/\partial N_{\bf 0}$.
It is convenient to introduce a dimensionless parameter $\eta
\left( N_{{\bf 0}},T\right) $ defined by

\begin{equation}
\eta \left( N_{{\bf 0}},T\right) =\frac{\mu \left( N_{{\bf 0}},T\right) }{
k_{B}T}.  \label{nu1}
\end{equation}
It can be
rewritten as:

\begin{equation}
\eta \left( N_{{\bf 0}},T\right) =\eta _{0}\left( 1-t^{2}\right) /t,
\label{nu}
\end{equation}
where $t$ is the reduced temperature introduced before and
 $\eta _{0}=\sqrt{2\zeta \left( 2\right) \theta }=1.62(a_s/l_{z})^{1/2}$
is a small parameter for the weakly interacting gas. Note that $\eta _{0}$
is irrelative to the number of the condensed atoms. For the low-dimensional
experiment \cite{EXPLOW}\ on Bose-condensed systems a simple calculation
gives $\eta _{0}=0.1$. We will see in the following that the role of
atom-atom interactions on the thermodynamic properties of the system is
determined by the parameter $\eta _{0}$.

For the quasi-2D interacting gas, with the lowest order approximation,
the occupation number $N_{{\bf n}}$
of the exited state ${\bf n\neq 0}$ is given by

\begin{equation}
N_{{\bf n}}={\frac{1}{\exp \left[ \beta {\left( n_{x}+n_{y}\right) \hbar
\omega _{\perp }}\right] \exp \lbrack \beta {n}_{z}{\hbar \omega _{z}-}\eta
\left( N_{{\bf 0}},T\right) \rbrack -1}.}  \label{exiteq2d}
\end{equation}
The term $\eta \left( N_{\bf 0},T\right) $ in the above equation represents
the interaction correction due to the interatomic interacion in the condensate.
Different from the ideal gas, we see from Eq. (\ref{exiteq2d}) that the
occurrence of the quasi-2D interacting gas should satisfy the condition ${
\hbar \omega _{z}>}\mu \left( N_{{\bf 0}},T\right) >>\hbar \omega _{\perp }$
(see also Ref. \cite{EXPLOW}). Note that the occurrence of a quasi-2D ideal gas
should satisfy the condition ${\hbar \omega _{z}>}k_{B}T>>\hbar \omega _{\perp }$.

Omitting the interaction between condensed and normal gases, from Eq. (\ref
{alphafunction}), we get

\[
{{N_{{\bf 0}}=N-\sum_{n_{x},n_{y}\neq 0}^{\infty }}\left[ {\frac{1}{\exp
\left[ \beta \left( n_{x}+n_{y}\right) \hbar \omega _{\perp }\right] {\exp }
\left[ -\eta \left( N_{{\bf 0}},T\right) \right] \exp \lbrack -{\alpha
\left( N,N_{{\bf {0}}}\right) }\rbrack -1}}\right] } 
\]

\begin{equation}
-{\sum_{n_{z}=1}^{\infty }\sum_{n_{x},n_{y}=0}^{\infty }}\left[ {\frac{1}{
\exp \left[ \beta \left( n_{x}+n_{y}\right) \hbar \omega _{\perp }\right]
\exp \left[ \beta n_{z}\hbar \omega _{z}-\eta \left( N_{{\bf 0}},T\right)
\right] \exp \lbrack -{\alpha \left( N,N_{{\bf {0}}}\right) }\rbrack -1}}
\right] .  \label{q2dit}
\end{equation}
By setting ${\alpha \left( N,N_{{\bf {0}}}\right) =0}$ in Eq. (\ref{q2dit}), 
${{N_{{\bf 0}}^{p}}}$ is given by

\[
{{N_{{\bf 0}}^{p}=N-\sum_{n_{x},n_{y}\neq 0}^{\infty }}\left[ {\frac{1}{\exp
\left[ \beta \left( n_{x}+n_{y}\right) \hbar \omega _{\perp }\right] {\exp }
\left[ -\eta \left( N_{{\bf 0}},T\right) \right] -1}}\right] } 
\]

\begin{equation}
~~~~~~~~~~~~~~~~~~~~~~-{\sum_{n_{z}=1}^{\infty }\sum_{n_{x},n_{y}=0}^{\infty
}}\left[ {\frac{1}{\exp \left[ \beta \left( n_{x}+n_{y}\right) \hbar \omega
_{\perp }\right] \exp \left[ \beta n_{z}\hbar \omega _{z}-\eta \left( N_{
{\bf 0}},T\right) \right] -1}}\right] .  \label{q2dipit}
\end{equation}

From Eqs. (\ref{q2dit}) and (\ref{q2dipit}), we get

\begin{equation}
{{N_{{\bf 0}}-N_{{\bf 0}}^{p}=\Upsilon }}_{2D}+\Upsilon _{mod},
\label{itnop}
\end{equation}
where

\[
{{\Upsilon }_{2D}=\sum_{n_{x},n_{y}\neq 0}^{\infty }\left[ {\frac{1}{\exp
\left[ \beta \left( n_{x}+n_{y}\right) \hbar \omega _{\perp }-\eta \left( N_{
{\bf 0}},T\right) \right] -1}}\right. } 
\]

\begin{equation}
\left.{-}\frac{1}{\exp \left[ \beta \left( n_{x}+n_{y}\right) \hbar \omega _{\perp
}-\eta \left( N_{{\bf 0}},T\right) \right] \exp \lbrack -{\alpha \left( N,N_{
{\bf {0}}}\right) }\rbrack -1}\right] ,  \label{gama2d}
\end{equation}
and

\[
\Upsilon _{mod}={\sum_{n_{z}=1}^{\infty }\sum_{n_{x},n_{y}=0}^{\infty }}
\left[\frac{1}{\exp \left[ \beta \left( n_{x}+n_{y}\right) \hbar \omega
_{\perp }\right] \exp \left[ \beta n_{z}\hbar \omega _{z}-\eta \left( N_{
{\bf 0} },T\right) \right] -1}\right. 
\]

\begin{equation}
\left. {-}\frac{1}{\exp \left[ \beta \left( n_{x}+n_{y}\right) \hbar \omega
_{\perp }\right] \exp \left[ \beta n_{z}\hbar \omega _{z}-\eta \left( N_{
{\bf 0}},T\right) \right] \exp \lbrack -{\alpha \left( N,N_{{\bf {0}}%
}\right) }\rbrack -1}\right] .  \label{gamamod}
\end{equation}

\subsection{Pure two-dimensional interacting Bose gases}

Firstly we discuss a purely 2D interacting gas where $\Upsilon _{mod}$ can
be omitted because $\beta \hbar \omega _{z}\rightarrow \infty $ in this
situation. Thus for the purely 2D interacting gas, from Eq. (\ref{itnop}), one
gets

\[
{\ N_{{\bf {0}}}-N_{{\bf {0}}}^{p}=\frac{Nt^{2}}{\zeta \left( 2\right) }
\left\{ \left[ \eta \left( N_{{\bf 0}}^{p},T\right) -\eta \left( N_{{\bf 0}
},T\right) \right] +\right. }
\]

\begin{equation}
~~~~~~~~~~~~~~~~~~~~~~~~~~~~ \left.\left[ \eta \left( N_{{\bf 0}},T\right)
\ln \eta \left( N_{{\bf 0}},T\right) -\eta \left( N_{{\bf 0}}^{p},T\right)
\ln \eta \left( N_{{\bf 0}}^{p},T\right) \right] +\alpha \ln \eta \left( N_{
{\bf 0}},T\right) \right\}.  \label{nop2}
\end{equation}
When obtaining Eq. (\ref{nop2}), we have used $\eta \left( N_{{\bf 0}
}^{p},T\right) <<1$ by assuming that the interaction is weak.
In addition, when
obtaining this formula we have used the assumption ${\alpha \left( N,N_{
{\bf {0}}}\right) <<}\eta \left( N_{{\bf 0}}^{p},T\right) $ which may not
hold in the case of an extremely weak interaction. However, we will see in
the following that this assumption is reasonable over a wide range of the
experimental parameters.

Using the following Taylor  expansion

\begin{equation}
\eta \left( N_{{\bf 0}},T\right) =\eta \left( N_{{\bf 0}}^{p},T\right) -
\frac{\eta _{0}}{2Nt\left( 1-t^{2}\right) ^{1/2}}\left( N_{{\bf 0}}^{p}-N_{
{\bf 0}}\right) ,  \label{nutaylor}
\end{equation}
one gets the result for ${\alpha \left( N,N_{{\bf {0}}}\right) }$:

\begin{equation}
{\alpha \left( N,N_{{\bf {0}}}\right) =-}\frac{1+\eta _{0}t\left| \ln \eta
\left( N_{{\bf 0}}^{p},T\right) \right| /\left[ {\small 2\zeta }\left(
2\right) \left( 1-t^{2}\right) ^{1/2}\right] }{Nt^{2}\left| \ln \eta \left(
N_{{\bf 0}}^{p},T\right) \right| /\zeta \left( 2\right) }\left( N_{{\bf 0}
}-N_{{\bf 0}}^{p}\right) .  \label{italpha}
\end{equation}
Below the critical temperature, ${\alpha \left( N,N_{{\bf {0}}}\right) }$
can be approximated as:

\begin{equation}
{\alpha \left( N,N_{{\bf {0}}}\right) =-}\frac{N_{{\bf 0}}-N_{{\bf 0}}^{p}}{
Nt^{2}\left| \ln \eta \left( N_{{\bf 0}}^{p},T\right) \right| /\zeta \left(
2\right) }.  \label{ialphab}
\end{equation}
We see that the assumption ${\alpha \left( N,N_{{\bf {0}}}\right) <<}\eta
\left( N_{{\bf 0}}^{p},T\right) $ holds in the present low-dimensional 
Bose condensed experiment, where $\eta _{0}=0.1$\cite{EXPLOW}.
The distribution
function of the condensate below the critical temperature is then

\begin{equation}
G_{2D}=A_{2D}\exp \left[ {-}\frac{\left( N_{{\bf 0}}-N_{{\bf 0}}^{p}\right)
^{2}}{2Nt^{2}\left| \ln \eta \left( N_{{\bf 0}}^{p},T\right) \right| /\zeta
\left( 2\right) }\right] ,  \label{inter-dis}
\end{equation}
where $A_{2D}$ is a normalization constant. In addition, the most
probability value $N_{{\bf 0}}^{p}$ is given by

\begin{equation}
N_{{\bf 0}}^{p}={N-}Nt^{2}-N\eta _{0}\left( 1-t^{2}\right) t\left[ 1-\ln
\left( \eta _{0}\left( 1-t^{2}\right) /t\right) \right] /\zeta \left(
2\right) -tN^{1/2}\ln N{.}  \label{2ditnp}
\end{equation}

Below the critical temperature, the particle number fluctuations of the
condensate in the purely 2D interacting gas is given by

\begin{equation}
{\left\langle \delta ^{2}N_{{\bf 0}}\right\rangle }_{2D}=\frac{Nt^{2}\left|
\ln \left[ \eta _{0}\left( 1-t^{2}\right) /t\right] \right| }{\zeta \left(
2\right) }.  \label{q2dfit}
\end{equation}
We see that interaction between atoms greatly changes the behavior of the
particle number fluctuations, in comparison with the case of a 3D Bose gas
(see Ref. \cite{XIONG2}). For small $\eta _{0}$ considered in the present
study, the interaction between atoms have an effect of reducing the
particle number fluctuations. We should note that Eq. (\ref{q2dfit}) do not
hold in the case of extremely low $\eta _{0}$ and when $\eta _{0}$ near $1$.

From the formulas (\ref{mean}), (\ref{fluc}), (\ref{inter-dis}) and (\ref
{2ditnp}) we can obtain the numerical result of the mean ground state
occupation number and particle number fluctuations of the condensate. Shown
in Fig. 3 is the numerical result of ${\left\langle N_{{\bf 0}}\right\rangle
/N}$ (dot-dashed line) for the purely 2D interacting gas with the interaction
parameter $\eta _{0}=0.1$. The numerical result of $\delta N_{{\bf 0}}$
(dot-dashed line) with $\eta _{0}=0.1$ is demonstrated in Fig. 4. We see from
Fig. 4 that the interaction between atoms will remarkably lower the
particle number fluctuations. With the decrease of the trapping frequency in
the $z$ direction, the particle number fluctuations will increase as a
result of weakening the confinement in the $z$ direction.

\begin{figure}[tb]
\psfig{figure=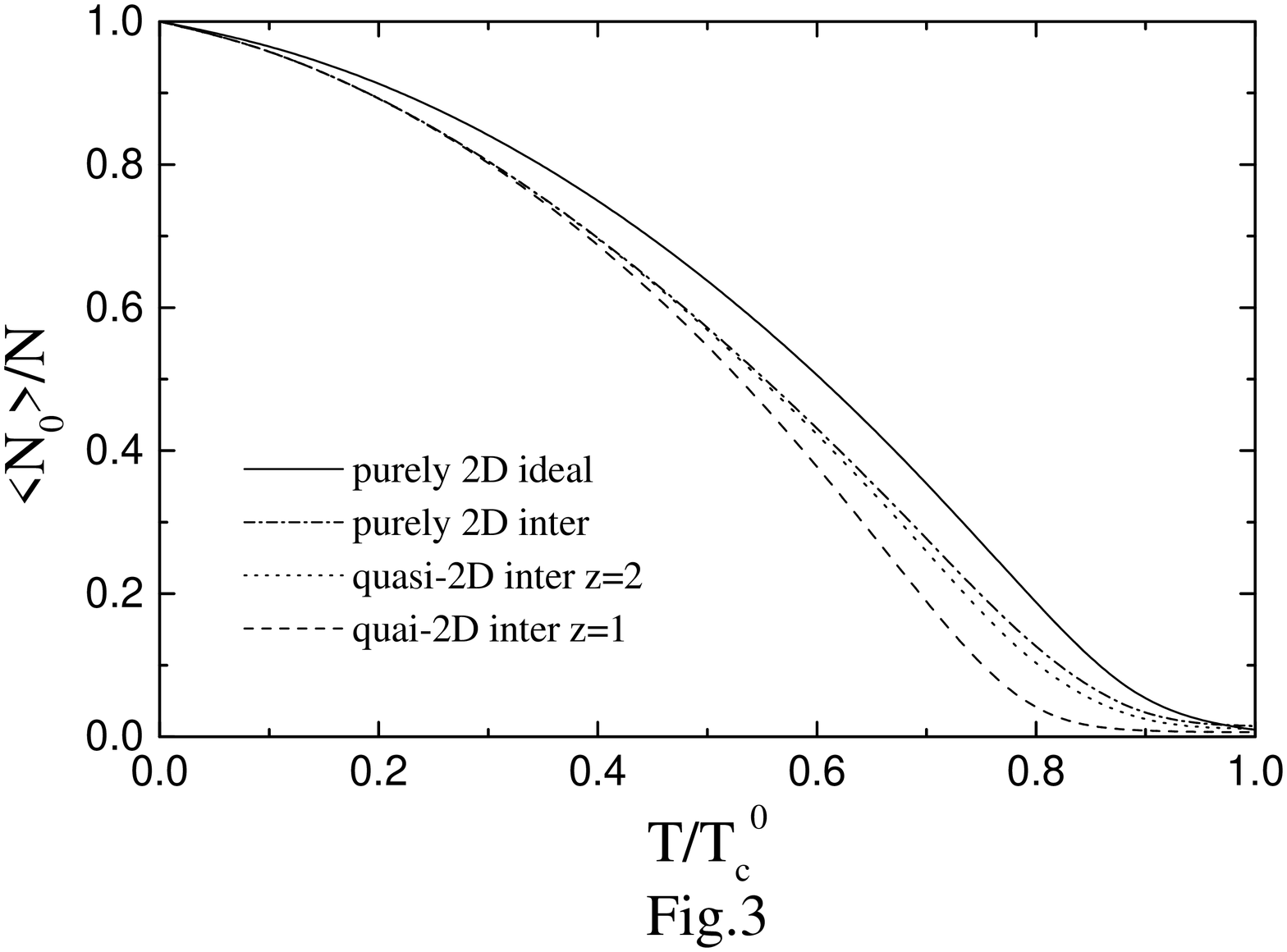,width=0.5\columnwidth,angle=270}
\caption{The dot-dashed line shows the numerical result of ${\left\langle N_{{\bf 0
}}\right\rangle /N}$ for $N=10^{3}$ purely 2D interacting bosons with $\eta
_{0}=0.1$, while the dashed and dotted lines display ${\left\langle N_{{\bf 0}%
}\right\rangle /N}$ of $N=10^{3}$ quasi-2D interacting bosons with 
$z=1$ and $z=2$, respectively. As illustrated in the
figure, the interaction between atoms has the effect of lowering the mean
ground state occupation number.
}
\end{figure}

\begin{figure}[tb]
\psfig{figure=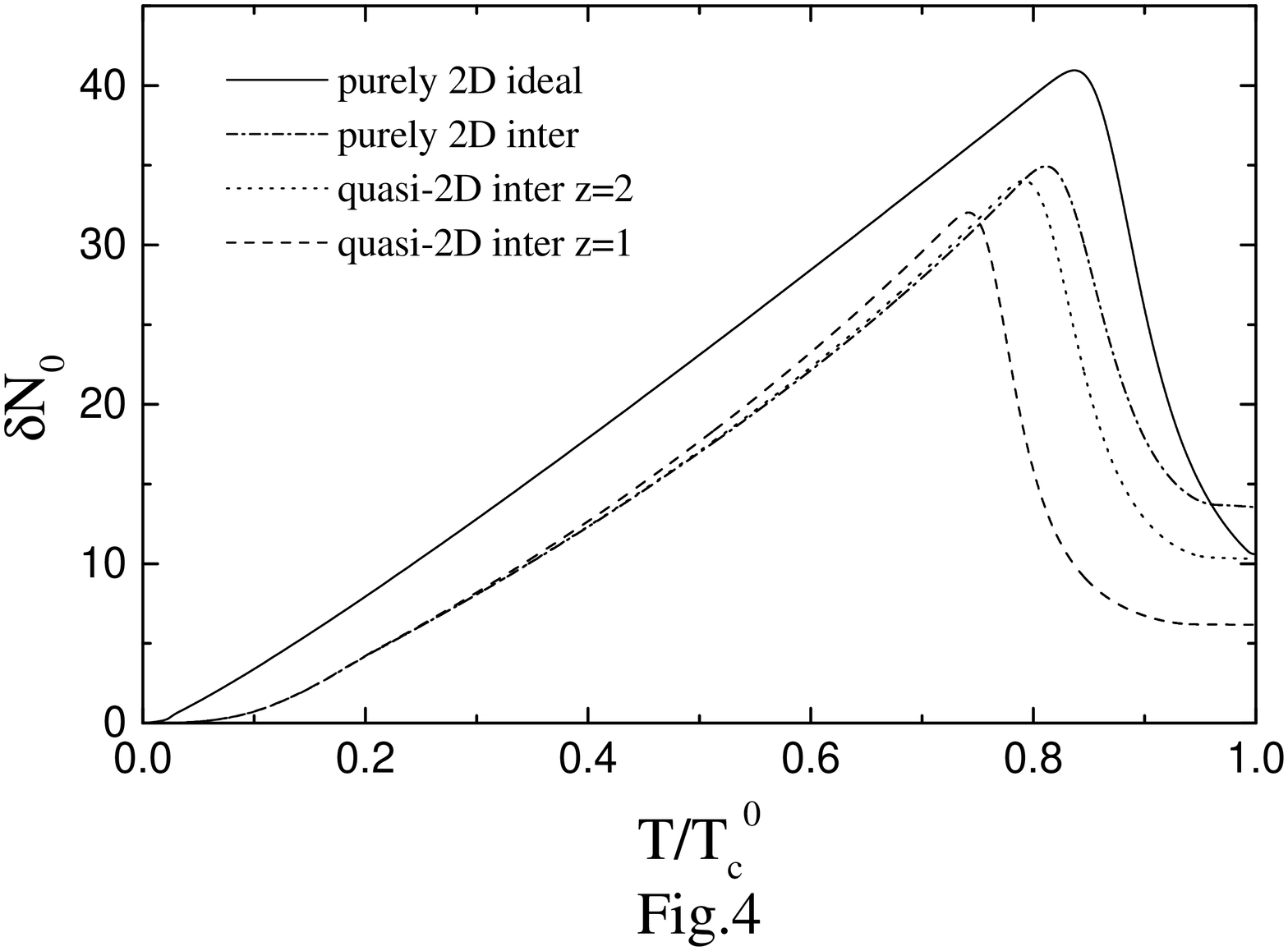,width=0.5\columnwidth,angle=270}
\caption{The dot-dashed line displays the numerical result of $\delta N_{{\bf 0}}$
for $N=10^{3}$ purely 2D interacting bosons with $\eta _{0}=0.1$, while the
dashed and dotted lines display $\delta N_{{\bf 0}}$ of $N=10^{3}$ quasi-2D interacting
bosons with $z=1$ and $z=2$, respectively. As
illustrated in the figure, below the citical temperature, the interaction between atoms has the effect of
reducing the particle number fluctuations of the condensate.
}
\end{figure}

\subsection{Quasi-two-dimensional interacting Bose gases}

Now we discuss the statistical properties of a quasi-2D interacting gas.
For the quasi-2D interacting gas, the second term $\Upsilon _{mod}$ on the
right hand side of Eq. (\ref{itnop}) can not be omitted. In the case of ${
\hbar \omega _{z}>}\mu \left( N_{{\bf 0}},T\right) $, $\Upsilon _{mod}$ can
be approximated as:

\begin{equation}
\Upsilon _{mod}=-{\alpha \left( N,N_{{\bf {0}}}\right) }\Theta _{2D},
\label{gammamod2}
\end{equation}
where $\Theta _{2D}$ is the constant given by Eq. (\ref{qtheta1}).
Using the approximation of $\Upsilon _{mod}$ it is easy to obtain the
probability distribution and particle number fluctuations of the quasi-2D
interacting gas below the critical temperature, which is given by

\begin{equation}
G_{q-2D}=A_{q-2D}\exp \left[ {-}\frac{\left( N_{{\bf 0}}-N_{{\bf 0}
}^{p}\right) ^{2}}{2Nt^{2}\left| \ln \eta \left( N_{{\bf 0}}^{p},T\right)
\right| /\zeta \left( 2\right) +2\Theta _{2D}}\right] ,  \label{q2dflucit}
\end{equation}

\begin{equation}
{\left\langle \delta ^{2}N_{{\bf 0}}\right\rangle }_{q-2D}=\frac{
Nt^{2}\left| \ln \left[ \eta _{0}\left( 1-t^{2}\right) /t\right] \right| }{
\zeta \left( 2\right) }+\Theta _{2D}.
\end{equation}
In addition, the most probability value is given by

\begin{equation}
N_{{\bf 0}}^{p}={N-}Nt^{2}-N\eta _{0}\left( 1-t^{2}\right) t\left[ 1-\ln
\left( \eta _{0}\left( 1-t^{2}\right) /t\right) \right] /\zeta \left(
2\right) -tN^{1/2}\ln N-\Theta _{2D}{.}  \label{q2dnop}
\end{equation}

Based on the formulas (\ref{mean}), (\ref{fluc}), (\ref{q2dflucit}), 
and (\ref{q2dnop}), the mean ground state
occupation number and particle number fluctuations of the condensate
are calculated numerically. In Fig. 3, the
numerical results of ${\left\langle N_{{\bf 0}}\right\rangle /N}$ 
for $N=10^{3}$ quasi-2D interacting bosons with $\eta _{0}=0.1$ 
are shown for $z=1$ (dashed line) and $z=2$ (dotted line), respectively. Shown in Fig. 4 is
the numerical result of $\delta N_{{\bf 0}}$ with the same interaction
parameter $\eta _{0}=0.1$. 
Similarly to 2D ideal gas, below the critical temperature,
we see that the particle number fluctuations in a quai-2D
interacting gas is slightly larger than the result of a purely 2D one.


\section{Discussion and Conclusion}


In this work, the statistical properties of pure- and quasi-2D Bose
gases are investigated within the canonical ensemble by using the 
method developed recently in Refs. \cite{XIONG1,XIONG2,XIONG3}. 
It is found that there is a
continuous crossover of the statistical properties from quasi-2D to purely
2D gases. Different from the case of a 3D gas, the effects of two-body
interaction and strong confinement in the $z$-direction change 
drastically the nature of the particle number
fluctuations for a quasi-2D Bose gas. 
For the quasi-2D interacting gas the role of atom-atom 
interaction on
the particle number fluctuations is determined by the 
reduced interaction parameter $\eta _{0}$, which
is determined not only by the $s$-wave scattering length but also
by the trapping frequency $\omega _{z}$ in 
the $z$ direction. Thus one can control the particle number fluctuations 
in the condensate 
by adjusting the trapping frequency in the $z$ direction. 
This unique property makes
it very promising to observe the particle number fluctuations 
in quasi-2D condensates.

In the present work, the contribution to the particle number fluctuations
due to collective excitations is omitted. For 3D Bose gas, as pointed out
in \cite{XIONG2}, the role of collective excitations would be very important
near zero temperature. By using Bogoliubov approach, the role of the collective
excitations on the condensate fluctuations would be very interested for the case of
2D Bose gas.
After the realization of low-dimensional Bose gas, the property of
collective excitations in 2D Bose gas becomes a very interested problem.
With the development of the theoretical research on the collective
excitations of 2D Bose condensed gas, we anticipate that the role of
the collective excitations on the condensate fluctuations would be given
in the near future.

It is obvious that the method given here can be also applied to other Bose
systems, such as an elongated (quasi-1D) trap. The discussion on the
statistical properties of the quasi-1D interacting Bose gas is beyond the scope
of this work and will be presented elsewhere.


\section*{Acknowledgments}


This work was supported by the Science Foundation of Zhijiang College,
Zhejiang University of Technology and Natural Science Foundation of China. 
G. X. Huang was supported by the National Natural Science
Foundation of China, the Trans-Century Training Programme Foundation for the
Talents and the University Key Teacher Foundation of Chinese Ministry of
Education. S. J. Liu and H. W. Xiong thank Professors Y. F Cao, G. S. Jia and J. F.
Shen for their enormous encouragement. In particular, the books presented by Professor
Y. F Cao is very useful for our research.

\end{document}